\documentclass[%
 reprint,
 superscriptaddress,
 amsmath,amssymb,
 aps,
 pra,
]{revtex4-2}

\usepackage[utf8]{inputenc}
\usepackage{graphicx}
\usepackage{dcolumn}
\usepackage{bm}
\usepackage{upgreek}
\usepackage{xcolor}
\usepackage{newunicodechar}
\newunicodechar{ }{\,}
\graphicspath{{./Figures/}}
\usepackage{hyperref}

\begin{document}
\title{Silicon Nitride Microresonator Raman Lasers}

\author{Yi Zheng}
\thanks{yizhen@dtu.dk}
\affiliation{DTU Electro, Department of Electrical and Photonic Engineering, Technical University of Denmark, 2800 Kongens Lyngby, Denmark}

\author{Haoyang Tan}
\affiliation{DTU Electro, Department of Electrical and Photonic Engineering, Technical University of Denmark, 2800 Kongens Lyngby, Denmark}

\author{Andreas Jacobsen}
\affiliation{DTU Electro, Department of Electrical and Photonic Engineering, Technical University of Denmark, 2800 Kongens Lyngby, Denmark}

\author{Yang Liu}
\affiliation{DTU Electro, Department of Electrical and Photonic Engineering, Technical University of Denmark, 2800 Kongens Lyngby, Denmark}

\author{Chaochao Ye}
\affiliation{DTU Electro, Department of Electrical and Photonic Engineering, Technical University of Denmark, 2800 Kongens Lyngby, Denmark}

\author{Yanjing Zhao}
\affiliation{DTU Electro, Department of Electrical and Photonic Engineering, Technical University of Denmark, 2800 Kongens Lyngby, Denmark}

\author{Cheng Xiang}
\affiliation{DTU Physics, Department of Physics, Technical University of Denmark, 2800 Kgs Lyngby, Denmark}

\author{Kresten Yvind}
\affiliation{DTU Electro, Department of Electrical and Photonic Engineering, Technical University of Denmark, 2800 Kongens Lyngby, Denmark}

\author{Minhao Pu}
\thanks{mipu@dtu.dk}
\affiliation{DTU Electro, Department of Electrical and Photonic Engineering, Technical University of Denmark, 2800 Kongens Lyngby, Denmark}

\date{\today}

\begin{abstract}
Silicon nitride (SiN) has emerged as a promising platform for integrated nonlinear photonics because of its low propagation loss, wide transparency window, and CMOS compatibility. Nonlinear processes arising from photon-electron interactions, such as Kerr frequency comb generation and second harmonic generation, have been extensively explored. In contrast, photon-phonon interaction-based nonlinearities, such as stimulated Raman scattering, remain largely unexplored in this integrated platform, despite their potential for broadband frequency conversion. Here, we demonstrate efficient Raman lasing in ultra-high-$Q$ SiN microresonators by harnessing the strong intracavity field enhancement and engineering the optical mode to overlap with the Raman-active silica cladding. Through dispersion engineering and waveguide geometry optimization, we suppress competing Kerr nonlinearities while enhancing Raman gain, achieving lasing with sub-2~$mW$ thresholds. We further investigate the trade-off between optical confinement and quality factor, revealing its impact on the overall nonlinear efficiency. Moreover, we also demonstrate broadband tunability of the Raman shift exceeding 120~$cm^{-1}$, enabled by the wide Raman gain spectrum of silica, offering new flexibility in designing integrated tunable Raman lasers. These results position SiN as a viable platform for chip-scale Raman lasers, expanding the nonlinear optics toolbox of the SiN platform and enabling compact, power-efficient light sources for applications in spectroscopy, optical communications, and quantum photonics.
\end{abstract}

\maketitle


\section{\label{sec:intro}Introduction}

The thin-film silicon nitride (SiN) platform has emerged as a highly versatile and scalable solution for integrated photonics, offering a unique combination of low optical loss, broad transparency spanning from the visible to the mid-infrared, and compatibility with CMOS fabrication processes. Traditionally, thin-film SiN has been primarily used as a low-loss, passive platform for high-quality factor ($Q$) resonators \cite{Blumenthal2018}, optical delay lines \cite{Bauters2011}, gyroscopes \cite{Gundavarapu2018}, and frequency filters \cite{Xiang2018} in integrated photonics. Beyond its role as a passive waveguide material, SiN plays a crucial role in hybrid photonic integration, enabling the cointegration of multiple active and nonlinear photonic functions by incorporating different materials onto a single chip \cite{Tran2022, Jin2021b}. While the hybrid integration approach has expanded the scope of SiN-based photonic devices, recent breakthroughs have demonstrated that the thin-film SiN platform itself can support new nonlinear optical processes previously considered unattainable. One major advancement is the realization of Kerr comb generation (KCG) in the thin-film SiN waveguide platform \cite{Yuan2023, Jin2021b}, despite its intrinsic normal dispersion, which initially prevented the Kerr parametric oscillation. Another major breakthrough is the realization of second-harmonic generation (SHG) in SiN \cite{Billat2017, Lu2021a}, which inherently forbids second-order nonlinear processes due to its amorphous structure. Stimulated Brillouin scattering (SBS) has also been demonstrated in the thin-film SiN platform \cite{Gundavarapu2019, Botter2022}. Although different nonlinear processes have been continuously unlocked in the thin-film SiN platform, stimulated Raman scattering (SRS) in SiN platform remains largely unexplored. Unlike KCG and SHG, which require stringent phase-matching conditions, SRS is mediated by material phonons, allowing efficient frequency shifts without phase-matching constraints, making it particularly attractive for broadband light generation and wavelength conversion. However, efficient SRS processes typically rely on material platforms with strong Raman gain coefficients, whereas SiN exhibits an intrinsically low Raman response, which has prevented the SiN platform from being used for Raman amplification and lasing.

\begin{figure*}
\includegraphics[scale=0.6]{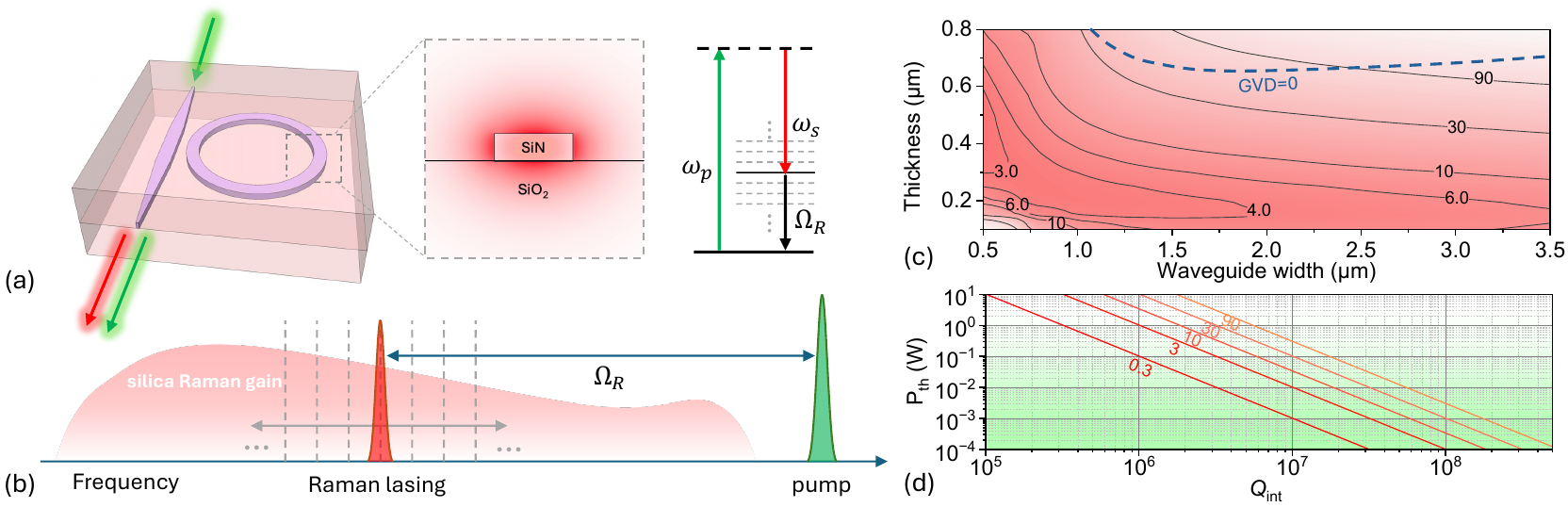}
\caption{\label{fig:principle}\textbf{Raman lasing in silicon nitride microresonators} (a) The pump light field can be concentrate in the silica cladding around a thin-film silicon nitride waveguide and resonantly enhanced in a microresonator, faciliating Raman lasing. The inset shows the mode profile of a SiN waveguide, emphasizing the overlap between the optical field and the silica cladding, with the waveguide core region highlighted in white. (b) Schematic of broadband Raman lasing enabled by the wide Raman gain bandwidth of silica. (c) Simulated effective $\mathrm{TE_{00}}$ mode area of silicon nitride waveguides, used to estimate the nonlinear Raman effects in the cladding material. Unit:~$\mu$m$^2$. The blue dashed line indicates the waveguide geometries with GVD=0. (d) Calculated Raman lasing threshold power as function of intrinsic $Q$ for microresonators with different $A_{eff}$.}
\end{figure*}
Raman lasers have been extensively studied in other CMOS-compatible materials such as silica and silicon, though each presents limitations. Silica-based whispering gallery mode (WGM) microresonators have demonstrated record-low Raman lasing thresholds \cite{Spillane2002b}, but their large footprint and integration challenges hinder scalability in photonic integrated circuits (PICs). While silica microresonators have been integrated with SiN circuits and KCG and Brillouin lasing have been demonstrated \cite{Yang2018}, to date, silica-based Raman lasing has not been realized in planar integrated platforms. Silicon-based Raman lasers were among the first explored in integrated photonics, leveraging silicon’s strong Raman scattering coefficient. Rong et al. pioneered the integrated silicon Raman laser \cite{Rong2005}, introducing lasing functionality to passive silicon photonics. Despite significant advancements \cite{Rong2008,Takahashi2013}, these devices are fundamentally limited by two-photon absorption (TPA) and free-carrier absorption (FCA), which introduce nonlinear losses at high pump powers in the telecom wavelength range. To mitigate these effects, techniques such as carrier extraction via reverse-biased p-i-n junctions have been employed \cite{Zhang2021, Zhang2022a}, though they introduce additional complexity in device design and fabrication.

Unlike silicon, SiN is free from TPA at telecom wavelengths and does not suffer from FCA, enabling high-power operation without nonlinear losses. Although SiN has a weak intrinsic Raman response, the ability to fabricate high-$Q$ microresonators on an integrated SiN platform enables strong nonlinear interactions, not only within the waveguide core but also in the waveguide cladding, which can serve as a Raman-active material. In this work, by leveraging ultra-high-$Q$ (tens of millions) microresonators and engineered waveguide light confinement, we enhance SRS within silica, a standard cladding material in CMOS fabrication. This enables efficient Raman lasing in thin-film SiN while suppressing competing KCG, achieving milliwatt-level lasing thresholds and broadband operation. Our results position thin-film SiN as a promising platform for chip-scale Raman lasers, addressing a critical gap in integrated nonlinear photonics and paving the way for future advancements in monolithic, multifunctional photonic systems.

\section{\label{sec:results}Results}
\textbf{Principle of Operation.} 
Figures.~\ref{fig:principle}(a,b) illustrate the operating principle of a SiN microresonator Raman laser. A pump light (green arrow) is coupled into a microring resonator embedded in silica (SiO$_2$), a Raman-active material, through a bus waveguide. The inset highlights how the optical mode (the fundamental TE mode: TE$_{00}$) is confined in a sub-micronmeter waveguide, where the overlap of the field with the surrounding silica enables SRS. In the microresonator, SRS is resonantly enhanced, leading to an efficient Raman lasing output (red arrow). The energy diagram on the right illustrates the Raman transition, where the pump frequency ($\omega_p$) is redshifted by the Raman shift ($\Omega_R$) to generate a new Stokes frequency ($\omega_s$). Figure.~\ref{fig:principle}(b) shows the spectral representation of Raman lasing, where the broad Raman gain of silica supports frequency conversion from the pump to multiple possible Raman lines, separated by the free spectral range (FSR) of the microresonator.   

To design a waveguide for efficient nonlinear interactions, we calculate the effective nonlinear interaction area $A_\text{{eff}}$ for SiN waveguides, considering that the pump and Stokes share the same spatial mode and the nonlinear interaction occurs in the waveguide cladding. Therefore, $A_{eff}$ defined here only takes into account the nonlinear interaction region outside the wavegide core, which is different from the conventional calculation for nonlinear processes that occur in the waveguide core. It is expressed as  \cite{Koos2007}:

\begin{equation}
    A_\text{{eff}} = \frac{Z_0^2}{n_\text{{cladding}}^2} 
    \frac{\left| \iint_\text{{all}} 
    \operatorname{Re} \left\{ \mathcal{E}(x,y) \times \mathcal{H}^{\star}(x,y) \right\} \cdot \mathbf{e}_z \,dx \,dy \right|^2}
    {\iint_\text{{cladding}} \left| \mathcal{E}(x,y) \right|^4 \,dx \,dy}
\end{equation}

\noindent where $Z_0$ is the free-space wave impedance, and $n_{\text{cladding}}$ is the refractive index of cladding.  $\mathcal{E}$ and $\mathcal{H}$ denote electric and magnetic field vectors of waveguide mode. $e_z$ is the unit vector along waveguide direction. Fig.~\ref{fig:principle}(c) shows the calculated $A_\text{{eff}}$ as a function of the thickness and width of the waveguide for the fundamental TE mode, where the color gradient represents the effective area; lower values indicate stronger mode confinement in the nonlinear cladding. A thin-film waveguide is preferred to enhance the evanescent field intensity in the nonlinear cladding. Although the minimum achievable $A_\text{{eff}}$ is larger compared to waveguides where the nonlinear interactions occur in the waveguide core, carefully optimized designs can reduce $A_\text{{eff}}$ to below 4~$\mu$m$^2$. This value is significantly smaller than that of a standard single-mode fiber (50-100~$\mu$m$^2$) and even smaller than that of highly nonlinear fibers (10-20 ~$\mu$m$^2$), which have been widely used in fiber-based Raman lasers. Furthermore,  $A_\text{{eff}}$ can remain below 15~$\mu$m$^2$ for waveguides thinner than 340~$nm$. The blue dashed line in Fig.~\ref{fig:principle}(c) denotes the zero-group velocity dispersion (GVD) boundary, with the upper region corresponding to anomalous-dispersion waveguide dimensions, which are essential for Kerr nonlinearities. Given the strong Kerr nonlinearity of SiN, it is also advantageous to employ thin-film SiN waveguides with normal dispersion and suppressed avoided mode crossing (AMX) \cite{Zhao2023} to inhibit nonlinear parametric oscillations and KCG, which would otherwise compete with Raman lasing \cite{Okawachi2017}.

\begin{figure*}[htbp]
\includegraphics[scale=0.68]{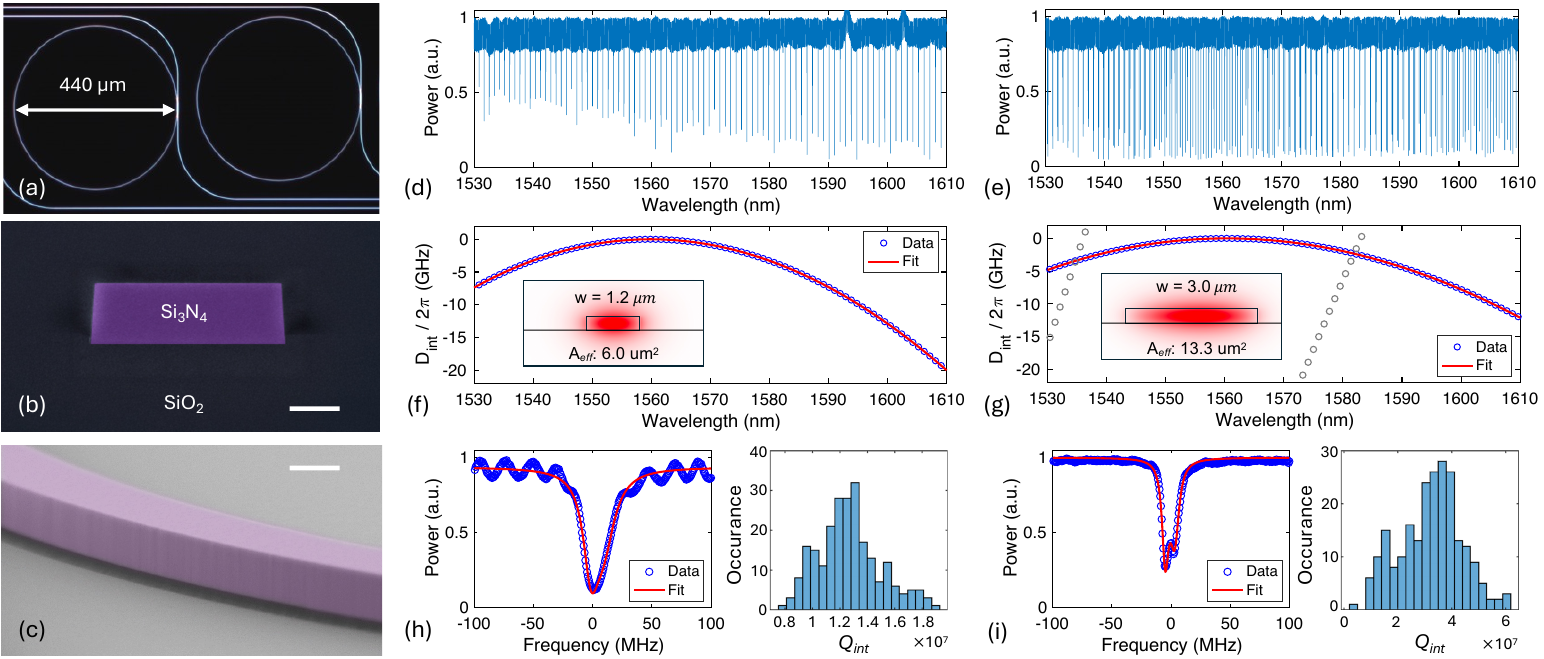}
\caption{\label{fig:resonator}\textbf{High-$Q$ silicon nitride microresonators} (a) Optical microscopy image of a SiN microring resonator with a 220~$\mu$m radius. (b) Scanning electron microscopy (SEM) image of the cross-section of a 340-$nm$-thick SiN waveguide cladded in SiO$_2$. (c) Bird-view SEM image of a SiN waveguide after the etching process. Scale bars, 300~$nm$ (b, c). (d,e) Measured transmission spectra of SiN microring resonators with different waveguide widths: (d) 1.2~$\mu$m and (e) 3~$\mu$m. (f,g) Measured integrated dispersion of the fundamental TE mode. Insets shows the corresponding field distribution for the fundamental TE mode. The grey circles are for the high-order TE mode. The normal dispersion and the absence of avoided mode crossing prevent Kerr comb generation from competing with the nonlinear Raman scattering. (h, i) (left) Measured typical transmission spectra at 1560.8~$nm$ (h) and 1567.5~$nm$ (i) exhibiting high $Q_\text{{int}}$: (h) 12.3 million and (i) 40.2 million. (right) Histrogram of measured $Q_{\text{int}}$ between 1560~$nm$ and 1630~$nm$ for microring resonators with different waveguide widths: (h) 1.2~$\mu$m and (i) 3~$\mu$m.}
\end{figure*}

The overall Raman response is influenced not only by light confinement in the cladding but also by the field enhancement in the microresonators. The Raman lasing threshold power ($P_\text{{th}}$) is described by following expression \cite{Spillane2002b, Kippenberg2004a}: 

\begin{equation}
    P_{\text{th}} = \frac{\pi^2 n_\text{{eff}}^2 A_\text{{eff}} L}{\lambda_P \lambda_R g_R} Q^P_c (\frac{1}{Q^P_l})^2 \frac{1}{Q^R_l}
\end{equation}\label{eq2}

\noindent where $n_{eff}$ is the effective index, $A_\text{{eff}}$ is the effective mode area of the pump mode, L is the cavity length, $\lambda_P$ and $\lambda_R$ are pump and Raman wavelengths, $g_R$ is the Raman gain coefficient, $Q^P_c$ is the coupling quality factor at pump wavelength, $Q^P_l$ and $Q^R_l$ are loaded quality factor at pump and Raman wavelengths, respectively.

Figure.~\ref{fig:principle}(d) shows the scaling of the Raman lasing threshold power ($P_\text{{th}}$) with the intrinsic quality factor ($Q_\text{{int}}$) for microresonators of varying  waveguide effective mode areas ($A_{eff}$). Assuming $Q^P_l=Q^R_l$ and critical-coupling for qualitative analysis, Eq. (2) simplifies to $P_{\text{th}} = \frac{8\pi^2 n_{\text{eff}}^2 A_{\text{eff}} L}{\lambda_P \lambda_R g_R} \frac{1}{Q^2_\text{{int}}}$. This reveals that $P_\text{{th}}$ scales linearly with $A_\text{{eff}}$ but inversely with $Q^2_\text{{int}}$. Since the light confinement and surface scattering loss (the dominating loss for SiN waveguides) are correlated with the waveguide dimension, the waveguide should be carefully designed to balance these factors for low-threshold operation. Narrowing a thin film waveguide helps reduce $A_\text{{eff}}$ and suppress high-order mode-induced AMX \cite{Ye2023RobustMicroresonators}, while widening the waveguide enhances the isolation of the fundamental TE mode from sidewall roughness, potentially leading to a higher $Q_\text{{int}}$. It is seen that milli-watt-level threshold Raman lasing can be realized with $A_\text{{eff}}$=10~$\mu$m$^2$ if $Q_{int}>2\times10^7$ can be obtained. Therefore, minimizing waveguide loss to achieve high $Q_\text{{int}}$ is essential for realizing efficient Raman lasing.\\

\textbf{Ultra-high-$Q$ silicon nitride microresonators}. We investigate thin-film SiN microring resonators (as shown in Fig.~\ref{fig:resonator}(a)) with varying waveguide widths to support both single-mode and multi-mode operations. Devices are fabricated by depositing a 340-$nm$-thick SiN film on a silicon substrate with a buffered oxide (BOX) layer using low-pressure chemical vapor deposition (LPCVD). Device patterns are defined via electron beam lithography (EBL) and transferred using inductively coupled plasma reactive ion etching (ICP-RIE) to form the waveguides. The structures are then clad in silicon dioxide (SiO$_2$). Fig.~\ref{fig:resonator}(b) presents a cross-sectional view of the waveguide, while Fig.~\ref{fig:resonator}(c) shows a side view of a SiN waveguide after the patterning process and prior to SiO$_2$ cladding deposition. Fig.~\ref{fig:resonator}(d) and (e) display typical transmission spectra for microring resonators with 1.2-$\mu$m- and 3-$\mu$m-wide waveguides, respectively, where the 1.2-$\mu$m design shows only the fundamental TE mode, while the 3-$\mu$m design exhibits two mode families (TE$_{00}$ and TE$_{10}$). From these transmission spectra, we extract both the dispersion characteristics and $Q_\text{{int}}$ of the microresonators. 

Although multimode waveguides can offer lower scattering loss than single-mode counterparts, unwanted coupling between the fundamental and high-order modes induced by fabrication or design-related imperfections can lead to AMX \cite{Kim2022}. Such mode interactions introduce resonance shifts and local dispersion variations, potentially triggering KCG \cite{Xue2015}, which compete with Raman lasing \cite{Okawachi2017}. To suppress AMX in multi-mode resonators, abrupt transition between straight and curved waveguides should be avoided. This can be achieved by using adiabatically tapered bends or adopting a fully circular ring cavity design \cite{Ye2022, Ji2022a, Ye2024}. In our implementation, we employ a circular ring resonator with a constant radius of 220~$\mu$m to avoid straight-to-bend transition mode mismatch and ensure the resonator device lies within a single EBL writing field, eliminating stitching-induced defects.

Figures~\ref{fig:resonator}(f, g) show the integrated dispersion (D$_{int}$) extracted from the 1.2-$\mu$m- and 3.0-$\mu$m-wide resonators. Both exhibit normal dispersion for the TE$_{00}$ mode and no AMX signatures, even in the multi-mode case where TE$_{10}$ crossings are observed around 1535~$nm$ and 1580~$nm$. The absence of AMX, combined with normal dispersion, ensures that KCG is effectively suppressed, allowing Raman scattering to dominate in both designs. 

\begin{figure*}
\includegraphics[scale=0.58]{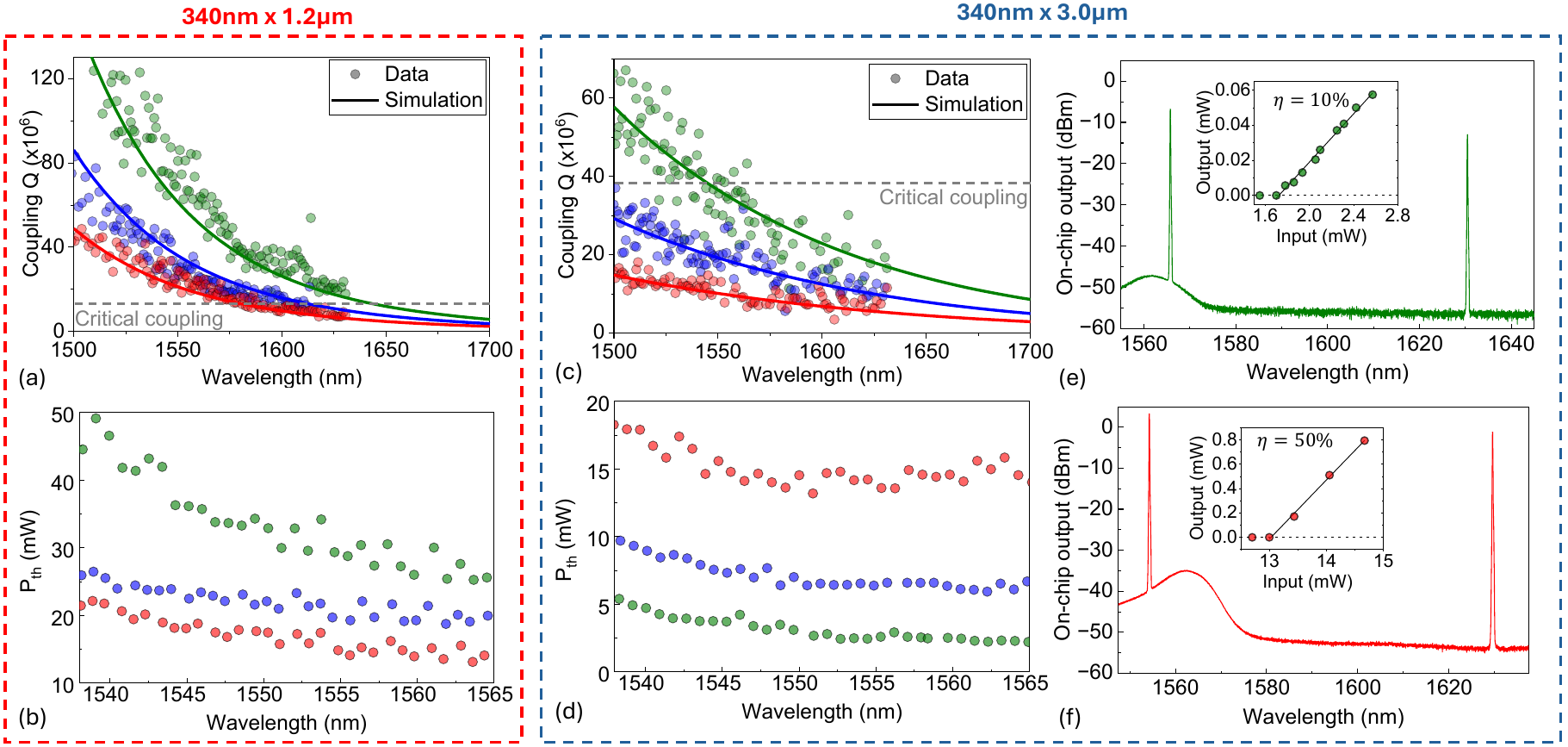}
\caption{\label{fig:Ramanlaser} \textbf{Raman lasing in SiN microresonators} (a), (c) Extracted (dots) and simulated (solid lines) coupling quality factor as a function of wavelength for microresonators with waveguide dimensions of 340~$nm$$\times$1.2~$\mu$m (a) and 340~$nm$$\times$3~$\mu$m (c). Dashed lines indicate the critical-coupling condition. Different colors represent different bus-to-ring coupling designs. (b), (d) Measured Raman lasing threshold power for pumping different resonances for the two microresonator geometries. Despite having a larger effective area, the multi-mode microresonators exhibit significantly higher $Q$, resulting in improved lasing performance with lower threshold power compared to the single-mode devices. (e, f) Measured Raman lasing spectra by pumping SiN microresonators (waveguide cross-sectional dimension: 340~$nm$$\times$3~$\mu$m) under different bus-to-ring coupling conditions: (e) critical-coupling and (f) over-coupling. Insets show the corresponding lasing threshold and slope efficiency. The critically coupled device exhibits the lowest threshold (1.8~$mW$), while the over-coupled device achieves a higher slope efficiency (50\%).}
\end{figure*}

The performance of the device in terms of $Q_\text{{int}}$ is summarized in Figs.~\ref{fig:resonator}(h, i). The left panels show representative measured transmission spectra with fitted resonance curves. Specifically, Fig~\ref{fig:resonator}(h)(left) displays a Lorentzian fit yielding a $Q_\text{{int}}$ of 1.2×10$^7$ at 1560.8~$nm$ for the 1.2-$\mu\mathrm{m}$-wide device, while Fig~\ref{fig:resonator}(i)(left) shows a split resonance with a fitted $Q_\text{{int}}$ of 4.0×10$^7$ at 1567.5~$nm$ for the 3-$\mu\mathrm{m}$-wide device. To evaluate the patterning process and assess the impact of waveguide geometry on device performance, we focus on the $Q_\text{{int}}$ data in the wavelength range from 1560~$nm$ to 1630~$nm$, thereby avoiding the N–H absorption band near 1510~$nm$ that can introduce excess optical loss \cite{Jin2021b}. The right panels present histograms of extracted $Q_\text{{int}}$ values, aggregated from three nominally identical devices for each waveguide width. 

The most probable $Q_\text{{int}}$ values (i.e., the peaks of the distributions) are 1.3×10$^7$ for the 1.2-$\mu\mathrm{m}$ design and 3.8×10$^7$ for the 3-$\mu\mathrm{m}$ design. The corresponding linear propagation losses, extracted from these most probable $Q_\text{{int}}$ values, are 2.63~$dB/m$ and 0.91~$dB/m$, respectively. These results indicate that increasing the width of the waveguide is an effective approach to improve $Q_\text{{int}}$, as the dominant loss mechanism remains scattering-induced. Narrower waveguides, such as the 1.2-$\mu\mathrm{m}$ design, are more susceptible to sidewall roughness due to stronger mode overlap with the waveguide boundaries. This contrast in mode confinement is illustrated by the simulated mode profiles in the insets of Figs.~\ref{fig:resonator}(f, g). Despite their distinct modal properties, both waveguide geometries achieve $Q_\text{{int}}$ exceeding ten million. This level of $Q_\text{{int}}$ values offers strong field enhancement critical for driving efficient Raman lasing.\\
 
\textbf{Efficient Raman lasing.}
To evaluate the device performance of the microresonators with both waveguide widths, we compared their Raman lasing threshold powers. The minimum threshold is expected under the critical coupling condition, which maximizes the intracavity pump power. To identify this point, we pumped the resonators under different coupling conditions (i.e., different $Q_c$ values) within the telecommunication C-band. Figs.~\ref{fig:Ramanlaser}(a, c) show the measured and simulated $Q_c$ as a function of wavelength for the 1.2-$\mu\mathrm{m}$- and 3-$\mu\mathrm{m}$-wide microresonators, respectively. Different colors represent different bus-to-ring coupling gaps, with red indicating the smallest gap and green indicating the largest gap. A decrease in $Q_c$ is observed at longer wavelengths due to weaker mode confinement. The dashed lines in Figs. \ref{fig:Ramanlaser}(a, c) denote the critical coupling condition. For both waveguide designs, we identified devices that achieve critical coupling within the C-band. Specifically, the red and green traces for the 1.2-$\mu\mathrm{m}$ and 3-$\mu\mathrm{m}$ wide devices, respectively, correspond to coupling gaps that enable critical coupling in this wavelength range. The lowest Raman threshold powers are obtained with the device closest to the critical coupling condition, corresponding to red and green colors for 1.2-$\mu\mathrm{m}$- and 3-$\mu\mathrm{m}$-wide MRRs. 

Figures~\ref{fig:Ramanlaser}(b, d) show the measured Raman lasing threshold power for devices with varying coupling conditions, corresponding to the $Q_c$ values in Figs.~\ref{fig:Ramanlaser}(a, c). The lowest threshold powers are measured as 13.1~$mW$ and 1.8~$mW$ for the 1.2-$\mu\mathrm{m}$- and 3-$\mu\mathrm{m}$-wide microresonators, respectively. As expected, the minimum threshold occurs near the critical coupling condition. Although the 1.2-$\mu\mathrm{m}$-wide devices have less than half of the effective mode area than the 3-$\mu\mathrm{m}$-wide counterparts, the significantly lower threshold observed in the 3-$\mu\mathrm{m}$-wide devices is attributed to their much higher $Q_\text{{int}}$. Since the Raman lasing threshold scales inversely with the $Q_{int}^2$, the more than threefold increase of $Q_\text{{int}}$ plays a crucial role in achieving low threshold power.

Figures~\ref{fig:Ramanlaser}(e,f) show the Raman spectra by pumping 3-$\mu\mathrm{m}$-wide microresonators under the critical- and over-coupling conditions, respectively. The insets display the measured output powers of the Raman signal versus pump power and the slope efficiencies are extracted. While the critical-coupled device exhibits the lowest threshold, the over-coupled device achieves a higher slope efficiency of 50\%, owing to more efficient out-coupling of the Raman signal from the resonator to the bus waveguide. The coupling condition also significantly impacts the achievable output power for single-mode Raman lasing. As shown in ~the inset of Fig.~\ref{fig:Ramanlaser}(f), the over-coupled device  delivers an output power approaching 1~$mW$, which is an order of magnitude higher than that from the critical-coupled device in Fig.~\ref{fig:Ramanlaser}(e).

\begin{figure*}
\includegraphics[scale=0.5]{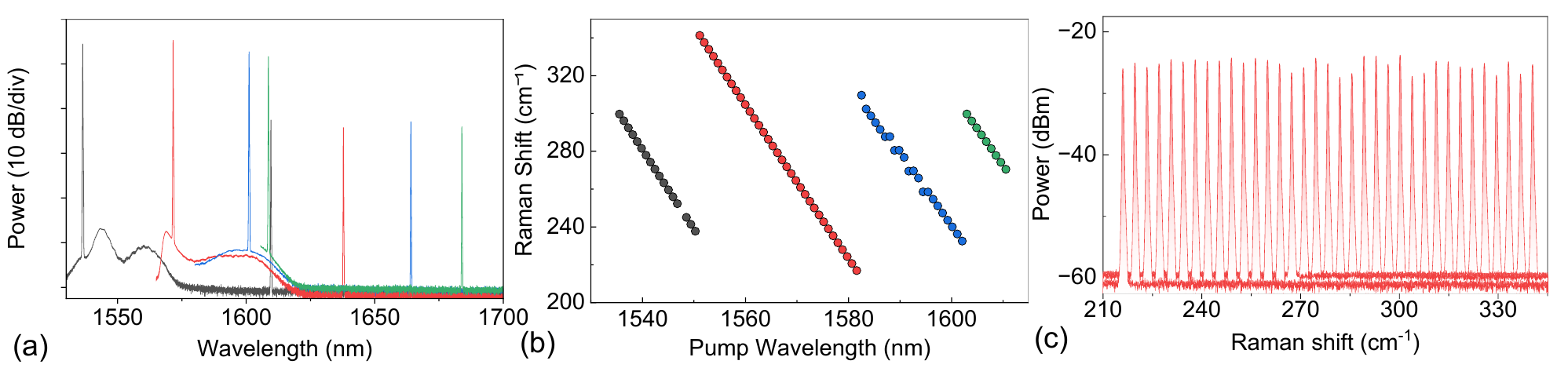}
\caption{\label{fig:tuning} \textbf{Raman laser tuning dynamics} (a) Measured Raman lasing spectra by pumping a single-mode microresonator (waveguide dimension: 340~$nm$ $\times$1.2~$\mu$m) at different wavelengths. (b) Measured Raman shift frequency as a function of the pump wavelength. Four distinct Raman frequencies are observed, with each of the Raman lasing frequencies remaining pinned within a specific pump tuning range. (c) Measured Raman lasing spectra as a function of Raman shift frequency. The broad bandwidth of the silica Raman gain enables Raman lasing over a large Raman shift frequency range.}
\end{figure*}

\textbf{Broadband Raman lasing operation}. 
Raman laser frequency tuning is typically achieved by adjusting the pump frequency since the Raman shifts in crystalline materials exhibit well-defined phonon vibrational frequencies. Consequently, the frequency-tuning range is fundamentally limited by the pump laser's tunability. In contrast, the Raman response observed in our system originates from silica, which is an amorphous material possessing a broad Raman gain spectrum. This property enables broadband frequency Raman shifts and enhanced wavelength-tuning flexibility. Figure.~\ref{fig:tuning}(a) shows the measured Raman lasing spectra from a  1.2-$\mu\mathrm{m}$-wide microresonator under wavelength-tuned pumping. While the Raman lasing wavelength can be tuned, the Raman shift frequency varied significantly. Figure~\ref {fig:tuning}(b) quantifies this behavior, showing Raman shift frequency spans from 217~$cm^{-1}$ to 341~$cm^{-1}$ with the pump tuning. 

Notably, the data points with a consistent color in Fig.~\ref {fig:tuning}(b) represent a fixed Raman lasing wavelength, which exhibits a pinning behavior over a specific pump range. For instance, the Raman laser wavelength was maintained at 1637.8~$nm$ while tuning the pump from 1555.1~$nm$ to 1581.6~$nm$. We attribute this wavelength pinning to the intracavity reflections, evidenced by the mode splittings shown in the transmission spectra. Strong intracavity reflections provide optical feedback that enhances SRS at specific resonances, thereby promoting single-mode lasing. This mechanism also enables tuning of the Raman shift frequency in our ultra-high-$Q$ SiN microresonators over 120~$cm^{-1}$, as further illustrated in Fig.~\ref{fig:tuning}(c), where the output spectra show stable single-mode Raman lasing throughout pump tuning. Leveraging the broad Raman gain bandwidth of silica, future implementations could combine the ultra-high-$Q$ SiN microresonators with tunable reflectors such as vernier resonators \cite{Tran2019TutorialIntegration}  to achieve Raman wavelength tuning while keeping the pump wavelength fixed. Furthermore, the broadband gain response relaxes the requirement for a small FSR typically imposed by the narrow Raman gain peak in crystalline materials. This allows the use of microresonators with small cavity volume (i.e., large FSR), which enhances field confinement and can improve Raman lasing efficiency.

\section{\label{sec:conclusion}Conclusion}

In summary, this work establishes thin-film SiN as a compelling alternative to silicon for CMOS-compatible integrated Raman lasers. Free from TPA at telecom wavelengths, the ultra-high-$Q$ SiN microresonators enable strong intracavity field buildup, resulting in low lasing thresholds and high output powers suitable for practical applications. The demonstrated milliwatt-level threshold and broadband operation pave the way for compact, chip-scale, and widely-tunable Raman lasers. Furthermore, the realization of SRS in ultra-high-$Q$ SiN microresonators expands the nonlinear optics toolkit available in this widely adopted platform, highlighting its versatility for exploring complex nonlinear interaction such as Raman–Kerr coupling \cite{Yang2017} and for enabling multifunction nonlinear photonic systems.\\

\textbf{Availability of data and materials}\\
The datasets generated and/or analyzed during in this study are available from the corresponding authors upon reasonable request.\\

\textbf{Competing interests}\\
The authors declare no competing financial interests.\\

\textbf{Funding}\\
This work is supported by European Research Council under the EU's Horizon 2020 research and innovation programme (grant agreement no. 853522, REFOCUS), Horizon Europe research and innovation programme under the Marie Skłodowska-Curie grant agreement No. 101119968 (MicrocombSys), Independent Research Fund Denmark (ifGREEN 3164-00307A), Danish National Research Foundation (SPOC ref. DNRF123), and Innovationsfonden (Green-COM 2079-00040B).\\

\textbf{Author contributions}\\
Y.Zheng and M.P. conceived the idea. Y.Zheng, H.T. and C.Y. designed the devices. Y.Zheng fabricated the devices. Y.Zheng, H.T., A.J., L.Y., C.X. and Y.Zhao performed the device charaterization. Y.Zheng, H.T, A.J., Y.L., C.Y., Y.Zhao, K.Y. and M.P. analyzed the data. K.Y. and M.P. supervised the work. Y.Zheng and M.P. wrote the manuscript with inputs from all authors.\\


 \bibliography{references}


\end{document}